\documentclass[journal=cgdefu,manuscript=article]{achemso}
\usepackage[version=3]{mhchem} 
\usepackage[T1]{fontenc}       
\usepackage{color}             

\author{Maciej \'{S}ciesiek}
\author{Wojciech Pacuski}
\author{Jean-Guy Rousset}
\affiliation{Institute of Experimental Physics, Faculty of Physics, University of Warsaw, Pasteura 5 St., 02-093 Warsaw, Poland}

\author{Magdalena Parli\'{n}ska-Wojtan}
\affiliation{Institute of Nuclear Physics, Polish Academy of Sciences, Radzikowskiego 152 St., 31-342 Krak\'{o}w, Poland}

\author{Andrzej Golnik}
\author{Jan Suffczy\'{n}ski}
\email{jan.suffczynski@fuw.edu.pl}
\affiliation{Institute of Experimental Physics, Faculty of Physics, University of Warsaw, Pasteura 5 St., 02-093 Warsaw, Poland}
\title[An \textsf{achemso} demo]
{Design and control of mode interaction in coupled ZnTe optical microcavities}

\keywords{coupled optical microcavities, Distributed Bragg Reflector, ZnTe, photonics}

\newcommand{\beginsupplement}{%
        \setcounter{table}{0}
        \renewcommand{\thetable}{S\arabic{table}}%
        \setcounter{figure}{0}
        \renewcommand{\thefigure}{S\arabic{figure}}%
     }

\begin{document}

\begin{tocentry}

\includegraphics[width = 83mm]{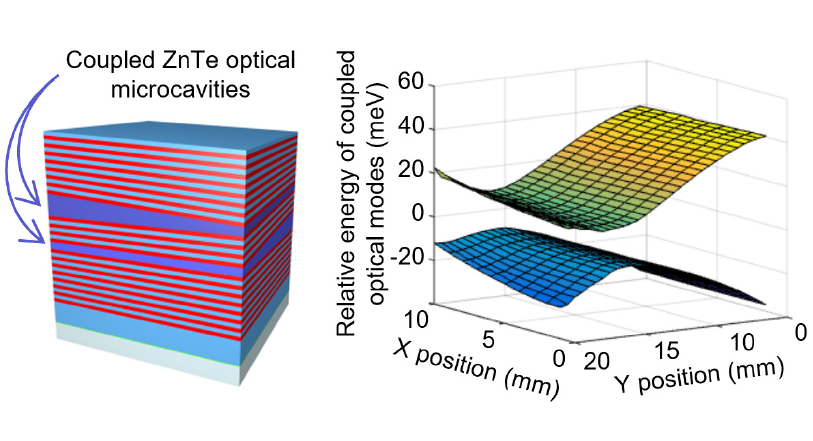}

\end{tocentry}

\begin{abstract}
The photonics involving II-VI epitaxial layers was limited so far to structures based on a single planar microcavity. Here, we present double, vertically coupled, ZnTe optical microcavities in planar and 3-D photonic molecule geometry. We design the structures with the help of transfer matrix method calculations and we establish their fabrication technology by molecular beam epitaxy. We characterize the samples by reflectivity spatial mapping and study them by measurements of angle-integrated and angle-resolved photoluminescence and reflectivity. We efficiently tailor the interaction strength of the cavities optical modes by an adjustment of the spatial separation between the microcavities, their thickness ratio and by the size of micropillars etched out of the planar structure. Coupling constants extracted from our measurements agree well with those determined in calculations in the frame of a tight-binding approach applied to one-dimensional photonic structures. 

\end{abstract}

\section{Introduction}
An evanescent field of the light confined in a semiconductor optical microcavity extends out of the microcavity boundaries. This effect allows for a coupling between optical modes of adjacent microcavities provided that their sizes, thus resonant frequencies, are close enough.\cite{Stanley:APL1994} The coupling brings a rich variety of phenomena, like an ultralow threshold optical parametric oscillation \cite{Diederichs:Nature2006} or a spontaneous mirror-symmetry breaking.\cite{Hamel:NaturePhot2015} It is also predicted to enable a generation of single photons under cw excitation thanks to a polariton blockade effect.\cite{Liew:PRL2010} Possible applications of coupled microcavities include areas of photonics, quantum information science and laser technology.\cite{Borishkina:Chapter2010}

The research devoted to coupled semiconductor microcavities has been so far limited mostly to the systems based on III-V compounds. At the origin a dominant direction of the investigations was the light-matter coupling in III-V planar structures coupled through a partially transparent Bragg reflector\cite{Stanley:APL1994, Armitage:PRB1998, Panzarini:PRB1999}. Next, the studies turned to {\it horizontally} 3-D coupled photonic molecules providing a stronger light confinement than the planar structures. They were realized in geometries such as: side-coupled micropillars etched out of a planar microcavity\cite{Bayer:PRL1998, Karl:OptExpress2007, deVasconcellos:APL2011}, neighbouring photonic crystal defect cavities\cite{Sato:NaturePhot2012, Haddadi:OptExpress2014} or microdisks side-coupled through an air gap.\cite{Ishii:APL2005}

In turn, studies of photonic systems involving II-VI compounds involved mostly single microcavities either in the planar\cite{Kelkar:PRB1995, LeSiDang:PRL1998, Saba:Nature2001, Martin:PRL2002} or the micropillar geometry.\cite{Robin:APL2005, Lohmeyer:APL2006, Jakubczyk:ACSNano2014} The only work reporting on coupled photonic molecules in II-VI based structures, authored by Sebald {\it et al.} (Ref.~[\cite{Sebald:OptExpr2011}]), was realized in the geometry of side-coupled micropillars etched out of a single ZnSe microcavity. However, a technology for the production of coupled microcavities based on other II-VI materials, such as ZnTe, CdTe or ZnO has not been established so far. 

The large exciton binding energy (in tens of meV range) in ZnTe and ZnO makes them a particularly good material for implementation in photonic devices operating at higher temperatures with respect to their III-V counterparts and should facilitate advanced experiments, like for example coupling of distant quantum emitters through a delocalized cavity mode.\cite{Benyoucef:PRB2008}. Several interesting phenomena resulting from the strong light-matter interactions in these materials have been already shown.\cite{Jakubczyk:JAP2013, Zamfirescu:PRB2002, Sturm:NJofPhys2011}

Here we present double, coupled, ZnTe based microcavities in planar (1-D) and single micropillar (3-D) geometry. We start with designing the structures using transfer matrix method (TMM) simulations and calculations in the frame of the tight-binding approximation\cite{Bayindir:PRL2000, Bayindir:JOA2001} applied to the modes of a 1-D photonic structure. In particular, we determine the coupling strength of the modes of neighbouring cavities as a function of the cavities separation, \textsl{i. e.}, the number of pairs forming the Distributed Bragg Reflector (DBR) inserted between them. Next, we epitaxially grow the structures, changing the cavities separation from sample to sample. The gradients of the cavities width are intentionally perpendicular to each other. Since the energy of the cavity resonant mode depends on the cavity width, this allows us to tune the relative detuning between the modes by selecting a place on the sample. In such a way, we are able to tune the cavities interaction strength, as predicted by our calculations  and verified in photoluminescence and reflectivity measurements. In particular, we confirm the increase of the cavity modes coupling strength with decreasing the number of DBRs pairs separating the cavities predicted by our simulations. Etching micropillars out of the planar sample enables us to study the impact of an increasing photon confinement on the modes energy and the coupling strength. In contrast to the work of Sebald {\it et al.} (Ref.~[\cite{Sebald:OptExpr2011}]) in our case the microcavities are coupled \emph{vertically}, \emph{i.e.}, in the direction perpendicular to the microcavities plane. The cylindrical symmetry of a single micropillar and the resulting decreased sidewall losses should allow for a higher photon collection efficiency with respect to previous double micro-pillar geometries.\cite{Bayer:PRL1998, Karl:OptExpress2007, deVasconcellos:APL2011, Sebald:OptExpr2011} 

As such, the present work provides a proof of concept for the \emph{vertically} coupled planar cavities and 3-D photonic molecules in single micropillar geometry based on II-VI semiconductor compounds, which are propitious for the realization of hybrid optoelectronic circuits involving photonic elements. Moreover, a large Huang-Rhys factor proper for the ZnTe\cite{Zhang:PRB2012} makes the studied system highly promising for implementation in an innovative laser cooling\cite{Pringsheim:ZPhysik1929} scheme involving the pumping through the optical mode of a double microcavity structure. The maximum efficiency of the scheme would be expected for the pump laser tuned to the lower energy mode in a region of the sample where the modes energy difference is one or two longitudinal optical (LO) phonon energy. In such a case an efficient anti-Stokes emission accompanied by an annihilation of the LO phonon would occur through the higher cavity mode, while the Stokes emission would be suppressed by the stopband.

\section{Sample design and epitaxial growth}
\subsection{Sample design}

The structures comprising two ZnTe $\lambda$-cavities embedded between DBR mirrors lattice matched to ZnTe are designed using TMM calculations. In the calculations we assume that the \emph{bottom} and \emph{top} mirrors are formed by 16 and 15 DBRs pairs, respectively. The role of high refractive index material in the DBRs is played by ZnTe layers, while a short period MgSe/ZnTe/MgTe/ZnTe superlattice serves as a low refractive index material.~\cite{Pacuski:APL2009} The wavelength dependent refractive indices of ZnTe, MgSe and MgTe are taken from Ref.~[\cite{Pacuski:APL2009}]. We expect that the use of only binary compounds should result in a more stable growth and a higher quality of the structure than it was in previous approaches to II-VI photonic structures, which involved ternary compounds\cite{LeSiDang:PRL1998, Peiris:JAP1999, Saba:Nature2001, Martin:PRL2002, Lohmeyer:APL2006, Sebald:OptExpr2011} or combined epitaxial growth with oxide deposition.\cite{Kelkar:PRB1995,Robin:APL2005}

We conduct two series of TMM simulations: (i) in the first of them, only the thickness of one of the cavities is varied, while all the other parameters, in particular the width of the second cavity, are fixed. This way we obtain the cavities coupling strength as a function of their relative detuning. We perform the calculations for a number $N_{DBR}$ of the DBR pairs separating the microcavities varied in the range from 3 to 12. (ii) In the second one, only $N_{DBR}$ is varied (in the range from 3 to 17), while all the other parameters are fixed. In particular, in this case we assume that both cavities have the same thickness, thus they are maximally coupled. In this way we obtain the cavities coupling strength as a function of their spatial separation. 

The sample design assumes that the coupled optical modes are detuned from the absorption edge of ZnTe (2.26 eV at room temperature). However, in order to emulate a realistic spectral linewidth, which is usually increased due to non-zero absorption of the light within the sample, we introduce a net imaginary contribution to a dielectric function of the ZnTe layers forming the cavities and the DBRs. For the sake of simplicity we assume that its value ($0.01$) is independent of the light wavelength.

\begin{figure}
  \includegraphics[width=0.6\linewidth]{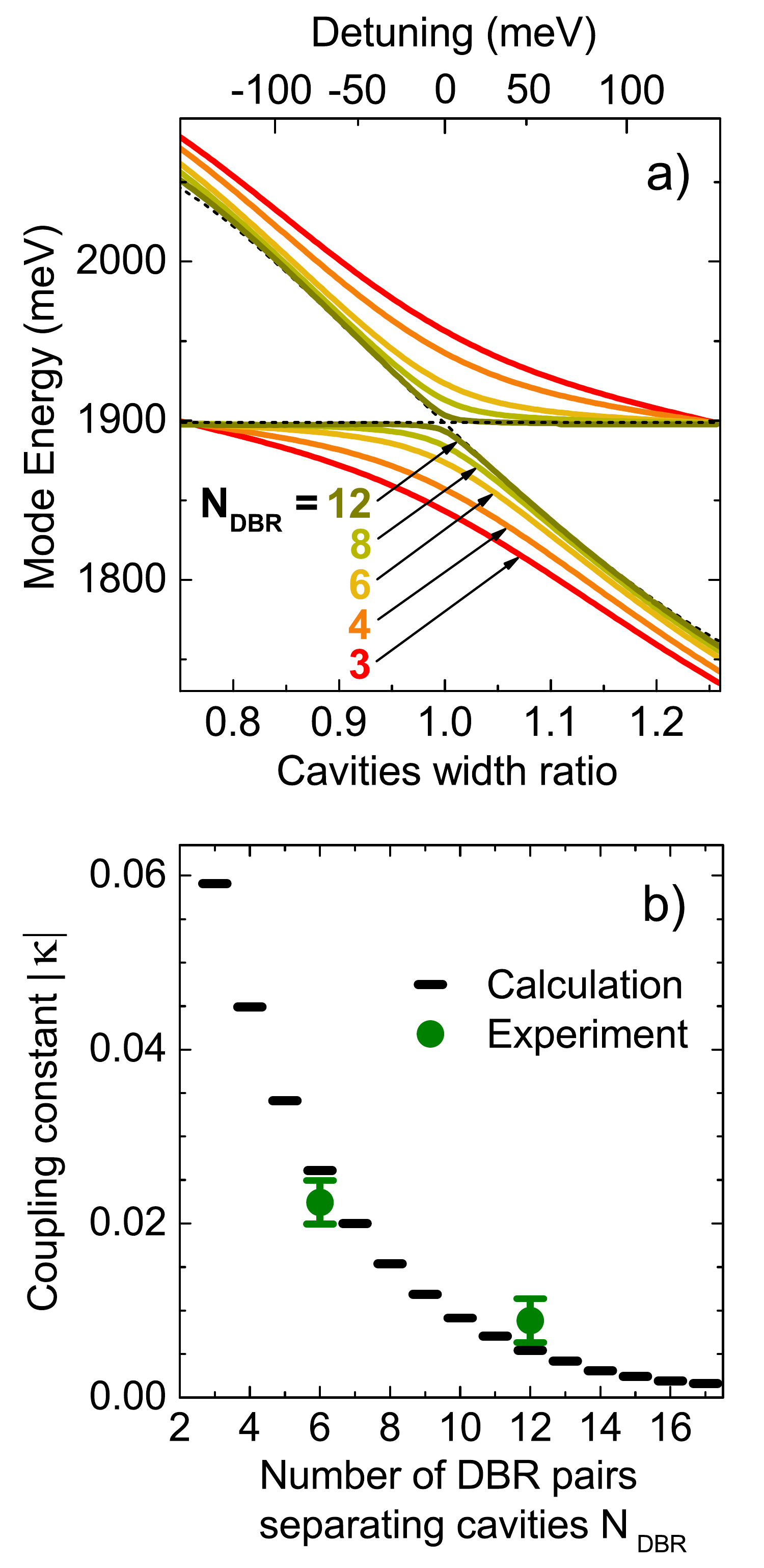}
\caption{a) Energy of the eigenmodes of two coupled ZnTe planar microcavities calculated as a function of the cavities ratio (bottom X axis) and the detuning between the modes of uncoupled microcavities (top X axis), for a number $N_{DBR}$ of the DBR pairs separating the cavities in the range from 3 to 12.
b) Constant |$\kappa$| characterizing the strength of the microcavities mode coupling calculated as a function of $N_{DBR}$.}
  \label{fig:Simulation}
\end{figure}

As a result of the TMM calculations we obtain reflectivity spectra of the structure. We determine the eigenmodes energy by fitting a sum of two Lorentzian curves to the two minima in the stopband region of the simulated spectra. The obtained dependence of the eigenmodes energy on the relative detuning  between the uncoupled modes (simulation (i)) is shown in Fig.~\ref{fig:Simulation}a) for an exemplary structure, where the fixed uncoupled cavity mode is centred at 650~nm (1910~meV).
When the detuning attains zero, a clear modes anticrossing is evidenced. The energy separation of the eigenmodes at the zero detuning increases when the $N_{DBR}$ decreases, which reflects the increasing modes interaction strength with a decreasing inter-cavity distance.

As shown by Bayindir and co-workers, a single dimensionless constant $\kappa$ derived from the eigenmodes energy difference at resonance is enough to characterize the coupling strength of 1-D microcavity modes.\cite{Bayindir:PRL2000, Bayindir:JOA2001} In the frame of a tight binding approximation applied to the optical modes localized in neighbouring microcavities $\kappa = \beta - \alpha$, where $\alpha$ and $\beta$ are determined from the following equation:\cite{Bayindir:PRL2000, Bayindir:JOA2001}
\begin{equation}
\omega_{1,2}^2 = \Omega^2 \left( 1 \pm \beta \right) / \left(1 \pm \alpha \right),
\end{equation}
with $\omega_{1}$ and $\omega_{2}$ the eigenmodes energy of the coupled cavities at zero detuning, while $\Omega$ is the energy of an uncoupled individual cavity mode. A plot of |$\kappa$| as a function of $N_{DBR}$ (simulation (ii)) is presented in Fig.~\ref{fig:Simulation}b).\cite{comment_kappa} It shows clearly that the cavities interaction strength decreases with the cavities separation, with an asymptote to zero for an infinite $N_{DBR}$. We note that the order of magnitude of the simulated $\kappa$ is in agreement with the values reported previously for 1-D coupled photonic structures.\cite{Bayindir:PRL2000,Bayindir:JOA2001} Based on the results of the simulations we epitaxially grow two structures (see the next subsection) adjusting their parameters so that the mode interaction at resonance is characterized by two significantly different values of |$\kappa$| equal to 0.0261 or 0.0054.

\subsection{Sample epitaxial growth}
The two structures are grown by molecular beam epitaxy on a 2-inch diameter GaAs substrate with a 1~$\mu$m thick ZnTe layer acting as a buffer. They are designed as described in the previous subsection (in consecutive order): 16 DBR pairs, the \emph{bottom} microcavity, a separating DBR, the \emph{top} microcavity, 15 DBRs pairs. The separation between the two ZnTe $\lambda$-microcavities is either 6 or 12 DBR pairs (|$\kappa$| of 0.0261 or 0.0054, respectively). The \emph{top} microcavity contains an active layer of CdTe quantum dots formed out of 3 monolayers of CdTe. The QDs are introduced to provide an emission source feeding the mode emission in photoluminescence studies.\cite{Suffczynski:PRL2009} 

\begin{figure}
  \includegraphics[width=1\linewidth]{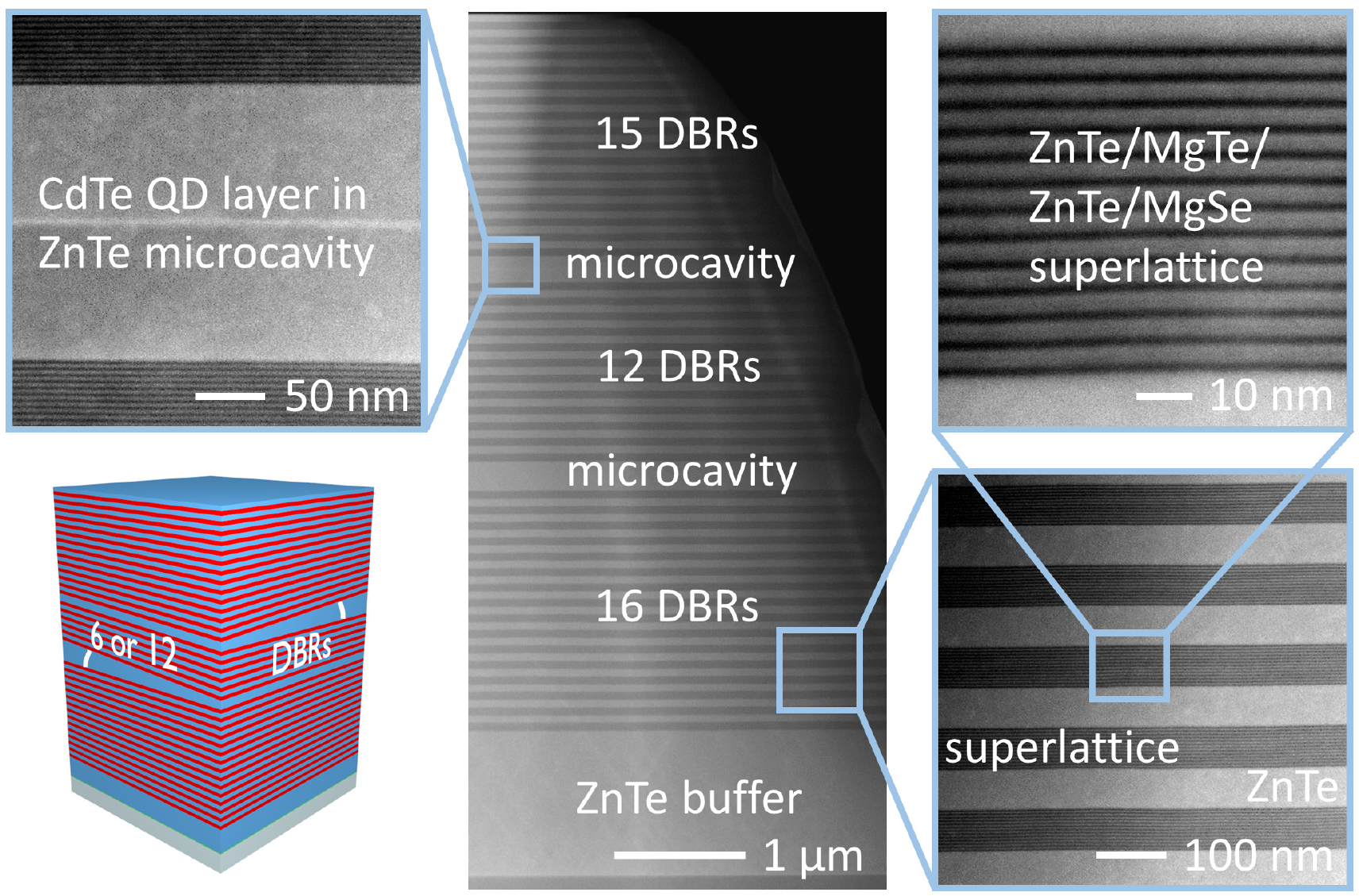}
\caption{A scheme showing the sample design comprising two vertically coupled ZnTe microcavities separated by either 6 or 12 DBR pairs with mutually perpendicular gradient of cavities thicknesses (lower-left corner). Scanning transmission electron microscopy images of the cross-section of the epitaxially grown sample (12 DBR pairs) with consecutive magnifications, as indicated by the respective scale bars.}
  \label{fig:Structures}
\end{figure}

We deliberately do not spin the substrate during the epitaxial growth, despite the spinning being usually used in order to avoid any gradient of thickness of the deposited layers resulting from a different position of different effusion cells with respect to the substrate. In this way we achieve that the thickness of the layers changes across the lateral position on the sample. For the growth of the top microcavity we rotate the substrate by 90 degrees and fix its position. Afterwards, we rotate the substrate back to the original orientation and we continue the growth. As a result of the assumed procedure, we obtain a mutually perpendicular gradient of the cavities thicknesses (see a schematic in Fig.~\ref{fig:Structures}). It enables us to continuously change the detuning between the modes of the cavities, hence the strength of the coupling between the microcavities, by adjusting the position at the sample surface. 
Note that within such a growth procedure the stopband spectral position also varies with the position at the sample surface, but for a given point it remains the same for the all three DBRs. 

The continuous loss of material volume in effusion cells during $\sim$~12 hours of the sample growth results in a decrease of the growth rate. We achieve a correct layers thickness by \emph{live} monitoring of the sample thickness during the growth by \emph{in situ} reflectivity and by adjusting the deposition time and/or the temperature of the sources, which controls the elements flow rate.

The result of scanning transmission electron microscopy characterization of the structure is shown in Figure~\ref{fig:Structures}. One can see a very good structural quality of the sample, with an absence of any dislocations. 
 
\section{Experiment}
The samples are studied by a reflectivity spatial mapping at room temperature performed in a custom, home made setup assuring an automated 7.5 cm per 7.5 cm 2-D movement with a resolution of 0.6 $\mu$m. The signal from the entire 2 inch wafer is collected with a step size of 1~mm. A halogen lamp light is focused to the spot diameter of around 0.1~mm at the sample surface.

Next, selected regions of the wafer are cleaved out into around 7 mm per 17 mm pieces. The cleaved sample is put onto the cold finger of a helium flow cryostat at a temperature adjusted in the 7~K - 300~K range. Micro-photoluminescence measurement is performed at the temperature of 50~K, which assures that optical transitions of individual quantum dots broaden and efficiently feed the optical modes of the cavities.\cite{Suffczynski:PRL2009} The signal from a 1 $\mu$m diameter spot is excited and collected through a microscope objective of numerical aperture NA = 0.75. The CW laser emitting at E$_{exc}$ = 2.33 eV ($\lambda_{exc}$ = 532 nm) serves as an excitation source. 

Angular resolution in microphotoluminescence and microreflectivity measurements is obtained by imaging the Fourier plane of the microscope objective on the slit of the spectrometer. To achieve this, an additional lens of focal length $f$ is inserted at a distance $f$ from the objective's Fourier plane. The lens imaging the signal on the spectrometer entrance is set at the position determined from the lens equation. The high numerical aperture of our objective enables a detection of photons emitted within a wide range of angles, from -50$^{\circ}$ to +50$^{\circ}$, without the need of any tilting of the sample.

Micropillars with diameters varying from 0.7 $\mu$m to 3 $\mu$m are etched out of the planar samples using a focused beam of Ga ions with an accelerating voltage up to 30 kV in a FEI Tecnai instrument. The procedure involves the etching of a large micropillar in the initial step and a subsequent decrease of its size to the desired diameter in subsequent steps. The ion current kept as low as 1.5~pA in the final polishing step assures a minimum roughness of the side surface of the micropillars. 

\section{Results and discussion}
\subsection{Coupled ZnTe planar optical microcavities}

\begin{figure}
\includegraphics[width=1\linewidth]{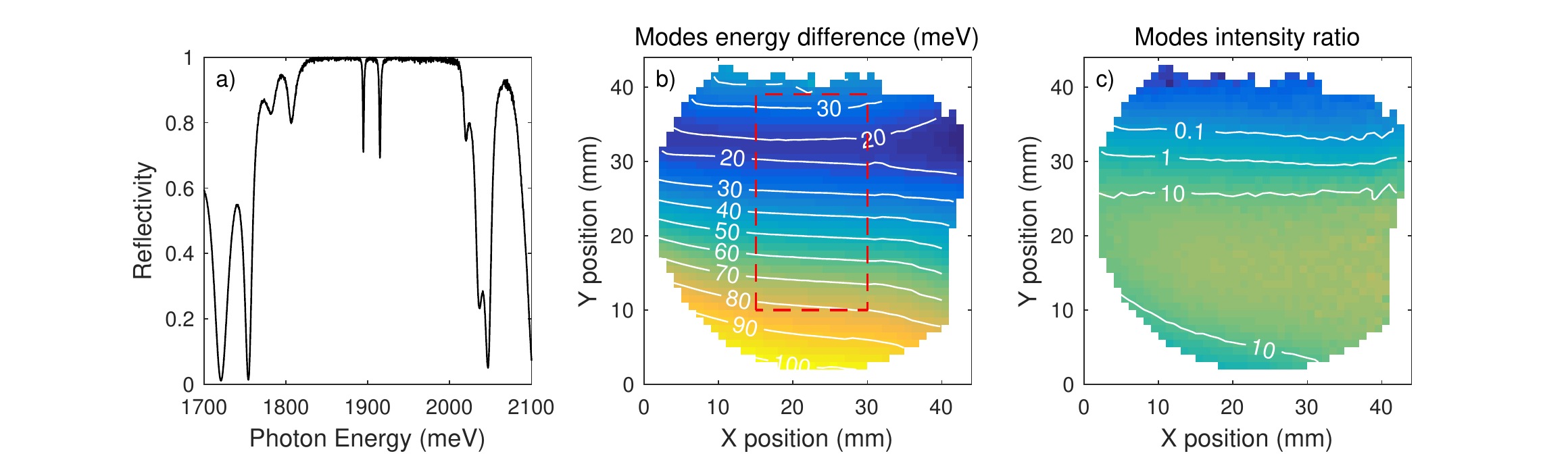}
\caption{a) An exemplary reflectivity spectrum of the 12 DBR sample at T = 300 K. b) Colour map showing the energy difference between the eigenmodes and c) the ratio of eigenmodes intensity determined for the whole area of the 12 DBR sample. The dashed line in panel b) highlights the region of the sample shown in enlargement in Fig.~4.
}
\label{fig:refl_mapping}
\end{figure}

An exemplary reflectivity spectrum of the 12 DBR sample is shown in Fig.~\ref{fig:refl_mapping}a). As predicted by the TMM calculations (see Suppl. Movie), it contains a stopband with two optical modes in its centre. The stopband is relatively wide spectrally (around 200 meV) and its reflectivity is above 99\%. This confirms that the layers involving binary compounds provide a reasonable contrast of refractive indices between the DBR layers and an overall good quality of the DBRs. 

In analogy to the case of electronic states localized in potential wells, the coupling of identical cavities leads to the emergence of spatially delocalized symmetric and antisymmetric coupled optical modes. As a result of the coupling they are split in energy. the presence of two extrema corresponding to the two modes is expected, in principle, for the optical spectrum of any photonic structure that contains two cavities. Thus, in order to prove the modes interaction one should plot the modes energy as a function of the modes detuning. The anticrossing behaviour of the modes is an unequivocal sign of the coupling in this case.

In order to prove the cavities coupling we perform a reflectivity mapping to plot the modes behaviour as a function of the detuning. We determine the modes energy and intensity by fitting with a sum of Lorentzian curves to the two minima in the stopband region of the spectra. The energy of the stopband centre is also determined (see Suppl. Fig.~\ref{fig:SIstopband_center}a)).

The energy difference between the eigenmodes and the ratio of the eigenmodes intensity for the 12 DBRs sample determined for the whole area of 2-inch sample are presented in Figs.~\ref{fig:refl_mapping}b) and ~\ref{fig:refl_mapping}c), respectively. As it can be seen, the ratio of the eigenmodes intensity is equal to unity for the region of the sample, where the energy difference between the modes is minimum. Such behaviour strongly suggests a mixing of the individual cavity modes resulting from their coupling. In regions of the sample, where the modes are strongly detuned, the energy and intensity of both modes vary independently of each other.

\begin{figure}
\includegraphics[width=1\linewidth]{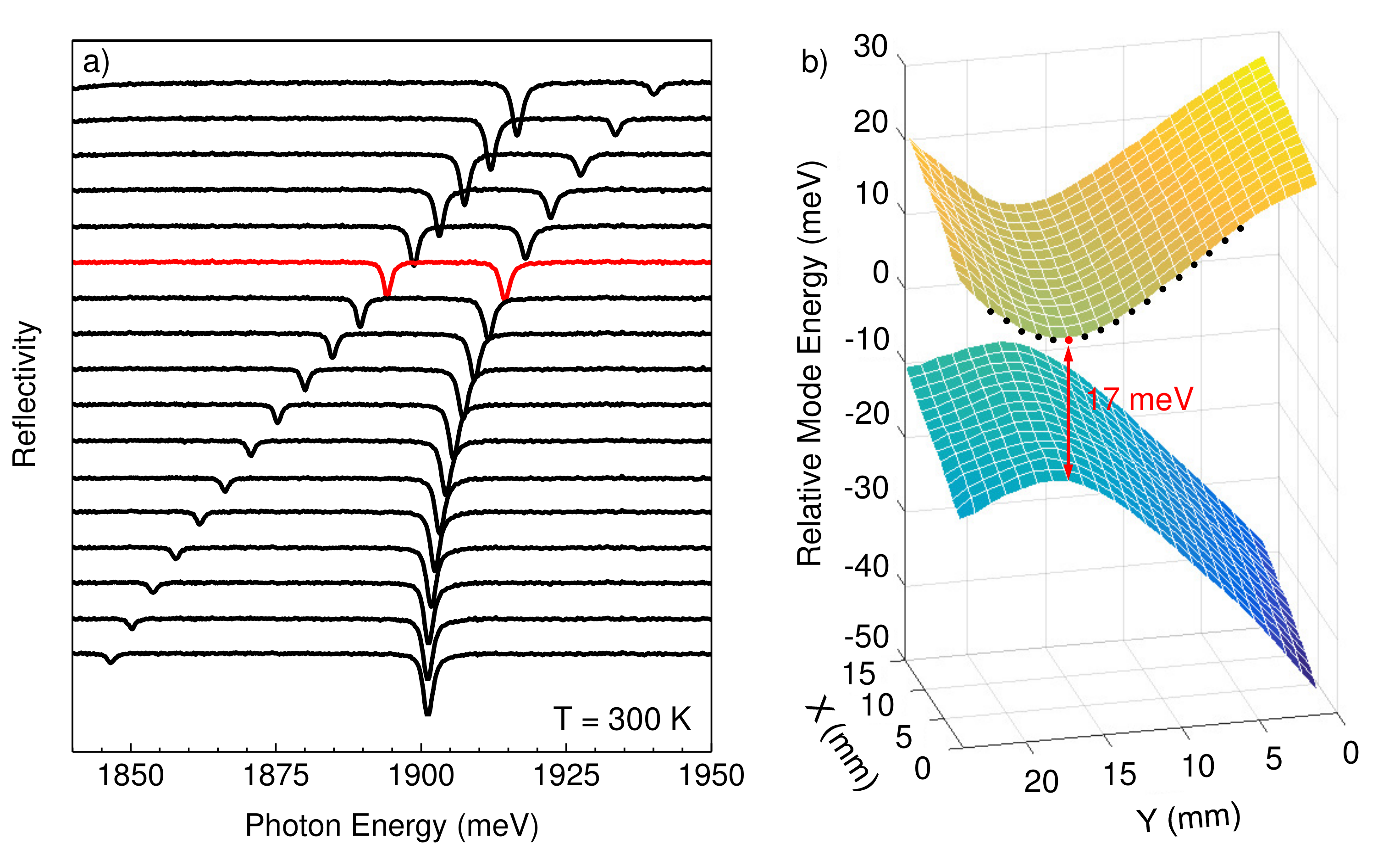}
\caption{a) Reflectivity spectra for consecutive points on the 12 DBR sample along the gradient width of the bottom microcavity. The reflectivity spectrum when the modes enter into resonance is indicated in red. b) Surface plot of the relative energies of the modes determined by the reflectivity mapping \emph{vs} the position on the sample. The dots indicate the positions, at which the reflectivity spectra shown in panel a) are taken. The surfaces interpolate the discrete values of the modes determined with 1~mm step in each dimension.
}
\label{fig:mode_energy}
\end{figure}

The reflectivity spectra for consecutive points selected every 1~mm along the thickness gradient of the bottom microcavity in the 12 DBRs sample are shown in Figure~\ref{fig:mode_energy}a) in the spectral range limited to the modes vicinity. With the change of the position on the sample, the mode initially at a lower energy (peaked at around 1850~meV at the bottom spectrum in Fig.~\ref{fig:mode_energy}a)) significantly shifts spectrally towards higher energy. The second mode initially keeps its position, however, when it is approached by the lower energy mode, a mutual pushing of the modes is evidenced. This clear anticrossing behaviour of the modes confirms that we truly accomplish the intended coupling between the two microcavities.

As a side remark, the mode initially at around 1850 meV is much weaker than the mode at around 1900 meV in the bottom curve of Fig.~\ref{fig:mode_energy}a). Its lowered intensity indicates that it is indeed linked to the bottom cavity, which signal is filtered by the stopband of the top cavity. Naturally, such attribution is possible only when the modes are far from resonance, since when the detuning approaches zero, their wavefunctions become mixed and spatially delocalized between the two microcavities.

In order to determine precisely the energy difference between the eigenmodes at resonance, we plot in Fig.~\ref{fig:mode_energy}b) the modes energy relatively to the stopband centre as a function of the position on the sample in the resonance region. The subtraction of the stopband energy allows us to eliminate a possible energy shift of the modes induced by the stopband shift resulting from the thickness gradient of the layers forming the DBRs. The eigenmodes splitting of 17 meV at resonance is found for the 12 DBRs sample. The splitting increases to 45 meV when the separation is decreased to 6 DBRs pairs (see Suppl. Figure~\ref{fig:17vs45}), in agreement with the expectation. The corresponding |$\kappa$| factors determined from the experiment amount to $0.0088 \pm 0.0025$ or $0.0220\pm 0.0025$, in the respective case of the 12 DBRs or 6 DBRs sample. As it is seen in Figure~\ref{fig:Simulation}, they remain in a good agreement with the predictions based on our calculations.

We perform angle resolved photoluminescence and reflectivity measurements, e.g., to determine the quality factor of the studied microcavities. The emission angle registered in the experiment encodes the in-plane component {\bf k$_{\parallel}$} of the momentum of the photon confined in the microcavity. The dependence between the photon energy and the {\bf k$_{\parallel}$} can be approximated by:\cite{AAmo:PhD2008} $E\left(k_{\parallel}\right) \sim \frac{c \, \hbar}{n_{\text{cav}}}\left(k_{\perp}+\frac{1}{2 k_{\perp}}k_{\parallel}^2\right)$, where $n_{\text{cav}}$ is the refractive index of the ZnTe microcavity. The $k_{\perp}$ being the photon momentum perpendicular to the microcavity plane is quantized due to the cavity resonant condition and takes a constant value for a given mode. Since the light rays leaving the microcavity at a given angle meet in a common point on the Fourier plane, a cross-section through the centre of the Fourier plane provides us with the dispersion relation E({\bf k$_{\parallel}$}). 

\begin{figure}
\includegraphics[width=0.65\linewidth]{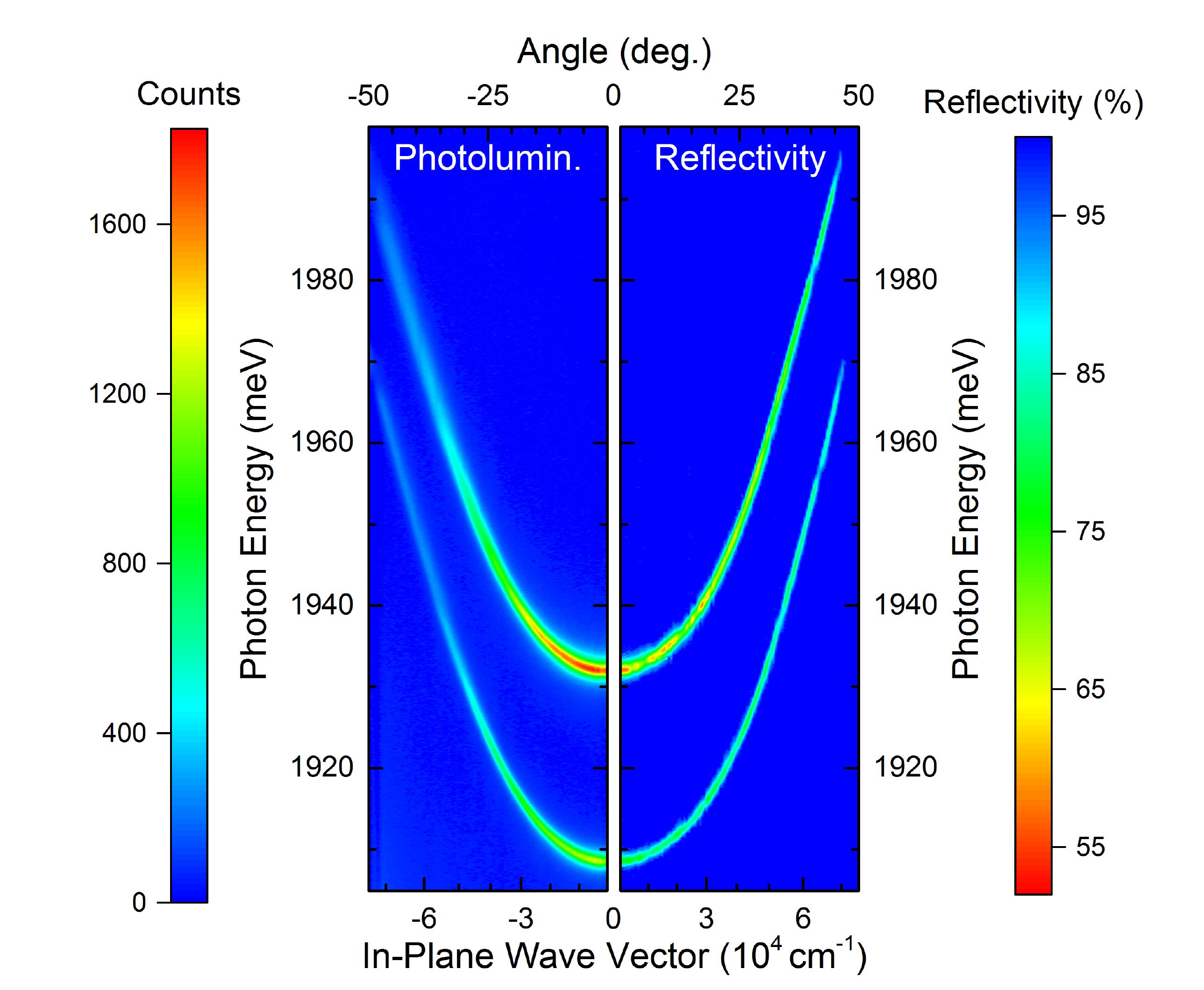}
\caption{Photoluminescence and reflectivity spectra of the 12 DBR sample resolved in photon energy and in plane momentum {\bf k$_{\parallel}$} for an exemplary point of the sample. T = 50 K.
}
\label{fig:kspace}
\end{figure}

The dispersion relation found for the 12 DBR sample is shown in Fig.~\ref{fig:kspace}. Let us note first that the E({\bf k$_{\parallel}$}) obtained from the reflectivity agrees well with the one coming from the photoluminescence measurement. It is also seen that the two extrema representing two eigenmodes of the coupled structure shift towards higher energy with an increasing photon in-plane wavevector. The E({\bf k$_{\parallel}$}) dependences of both eigenmodes take approximately a parabolic shape, similarly to what was previously observed in the case of microcavities confining a single mode.\cite{Kelkar:PRB1995, Armitage:PRB1998, Panzarini:PRB1999} The images shown in Figure~\ref{fig:kspace} are acquired for the point on the sample, where the modes are close to resonance. We have checked, however, that the observed E({\bf k$_{\parallel}$}) dependences are not sensitive to the modes relative detuning. 

The linewidth $\Delta E$ of the cavity resonance, that is the linewidth of the dispersion relation at {\bf k$_{\parallel}$ = 0} provides a quality factor of the microcavity Q = $E / \Delta E$. The linewidths of the modes at {\bf k$_{\parallel}$} = 0 are determined from the reflectivity measurement to 1.2~meV and 1.5~meV for the lower and higher energy mode, respectively. This yields a quality factor Q of the cavities in the range of 1600-1250. We link the slightly larger width found for the higher energy mode with the fact that this mode represents an "antibonding" photon state. In that case, the broadening may result from a higher degree of the delocalized photon penetration in the barrier. A similar broadening was previously observed for excitons confined at excited levels of a semiconductor quantum well.\cite{Dingle:PRL1974}  

The results presented in this Section show that the assumed sample design allows one to efficiently control the interaction between the modes of epitaxially grown, coupled ZnTe optical microcavities exhibiting a good crystal quality. The coarse adjustment is realized by the variation of the number of DBRs pairs separating the cavities. The fine control is obtained by selecting a position at the surface of the sample grown with a perpendicular gradient of the cavities thickness. 

\subsection{Coupled 3-D photonic molecules in single micropillar geometry}

A microstructuration of a planar cavity sample provides a mean for an additional confinement of the light. In structures embedding quantum emitters it allows therefore enhancing of the light-matter interaction constants resulting from a decreased mode volume.\cite{Yoshie:Nature2004, Reithmaier:Nature2004} In the case of the studied structures one should expect that the microstructuration will result in an increased interaction strength between the coupled modes resulting from a stronger radial localization of the photon.

\begin{figure}
  \includegraphics[width=1\linewidth]{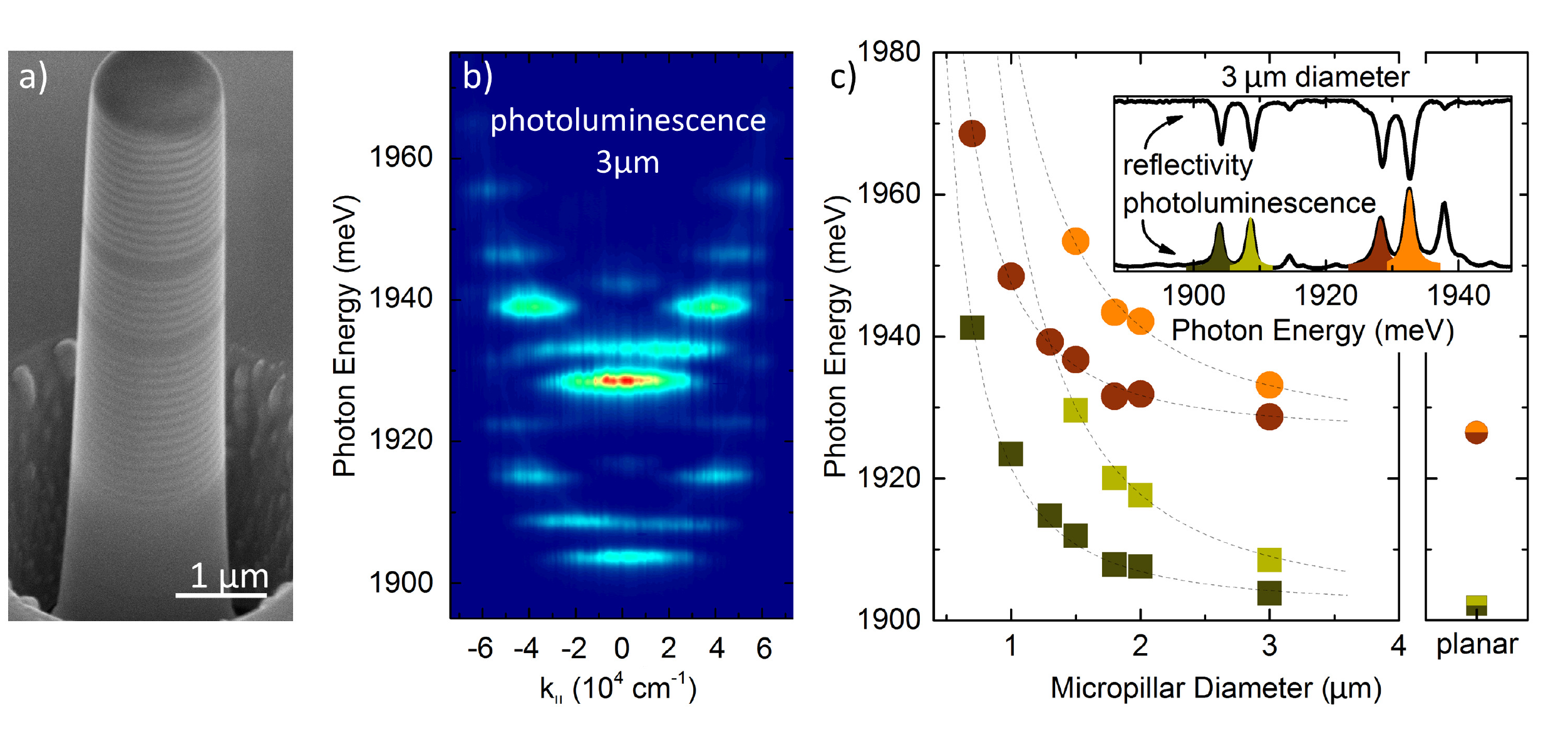}
\caption{a) Scanning electron microscopy image of an exemplary micropillar etched out of the 6 DBR sample. The 1~$\mu$m scal bar is indicated. b) Emission of a 3~$\mu$m micropillar from the 12 DBR sample as a function of the photon energy and the in-plane momentum, which reveals discrete submodes of the coupled 3-D photonic molecule. c) Energy dependence of the first two submodes on the micropillar diameter for the 12 DBR sample. Inset: exemplary reflectivity and photoluminescence spectra measured without angular resolution, along with fitted Lorentzian curves.}
  \label{fig:micropillars}
\end{figure}

Micropillars are etched out of the coupled planar microcavities (see Fig.~\ref{fig:micropillars}a) and Suppl. Movie) in a region where the microcavities are maximally coupled. Their diameter ranges from 3 $\mu$m down to 0.7 $\mu$m. Since a discontinuity of the refractive index at the micropillar sidewalls provides a photon confinement in the radial direction, the obtained coupled photonic molecule has a three-dimensional character. In previous approaches involving DBRs based micropillar microcavities\cite{Bayer:PRL1998, Karl:OptExpress2007, deVasconcellos:APL2011, Sebald:OptExpr2011} the photonic molecules were realized in the side-coupled micropillar geometry, where the cavities were \emph{horizontally} coupled. In contrast, the presented 3-D photonic molecule is realized in a single pillar geometry and with a \emph{vertical} coupling of the cavities.

The study of the molecules involves reflectivity and photoluminescence measured at T = 50 K. A decrease of the light spot down to 1~$\mu$m diameter enables us to address and collect the signal from individual micropillars. The microstructuration results in a quantization of the cavity mode into a set of discrete submodes (see Fig.~\ref{fig:micropillars}b and inset to Fig.~\ref{fig:micropillars}c). The highest energy submodes originating from the lower energy eigenmode spectrally overlap with the lowest energy submodes originating from the higher energy eigenmode. The photoluminescence measurement performed with angular resolution enables us to properly determine the origin and energy of a given submode. 

The energies of the first two submodes emerging from each eigenmode of the planar structure are determined by fitting with Lorentzian curve (see the inset to Fig.~\ref{fig:micropillars}c)). It is evidenced that with decreasing the pillar diameter, both the submodes energy and energy spacing between the consecutive submodes increase (see Fig.~\ref{fig:micropillars}c). This demonstrates an increased coupling strength of the modes resulting from the increasing photon confinement. The observed shape of the dependences of the submodes energy on the pillar diameter is such as observed previously for the modes of individual micropillars etched out of a single planar microcavity.\cite{Reithmaier:PRL1997, Lohmeyer:APL2006} 

This Section indicates that the sample microstructuration is an effective tool for tuning of the coupled modes energy, as well as for a control of the modes coupling strength. Thanks to a a high symmetry and resulting low sidewall losses, the presented innovative design of coupled 3-D photonic molecules in single micropillar geometry is promising for practical applications.
\section{Conclusions}

In the present work, we provide a proof of concept for the double coupled optical microcavities and the coupled photonic molecules in a single micropillar geometry based on ZnTe, being a representative member of II-VI compounds family. The design, growth and results of spectroscopy measurements confirming the interaction and control of the microcavities optical modes are presented for two samples differing by the microcavities separation.

The ability for tuning of the energy between the eigenmodes to one or two LO phonon energy in ZnTe ($\sim$ 26 meV)\cite{Nahory:PR1967} in 12 DBR od 6 DBR structure, respectively, confirms the potential of the presented structures for the application in the laser cooling scheme mentioned in the introductory section. A high exciton oscillator strength makes the studied system suitable also for achieving the interaction between distant quantum emitters such as quantum dots or quantum wells mediated by the delocalized optical mode. Thanks to the photon contribution to the wavefunction of the exciton and resulting emergence of polaritons,\cite{Kelkar:PRB1995,LeSiDang:PRL1998,Armitage:PRB1998,Panzarini:PRB1999} the exciton tunneling between the quantum wells embedded in two microcavites should be made possible. The tunneling in such a case would take place for a distance of the order of a micrometer, which exceeds for more than an order of magnitude the distances reported on previously for quantum well excitons.\cite{Lawrence:PRL1994} Moreover, the performance of the studied system could be further enhanced through the use of semimagnetic quantum wells or quantum dots embedding individual magnetic ions.\cite{Pacuski:CGDesign2014}

\section{Acknowledgements}

An access to the FEI Tecnai Osiris TEM instrument located at the Facility for Electron Microscopy \& Sample Preparation of the University of Rzesz\'{o}w, as well as a financial support of Polish National Science Centre (NCN) under the projects no. UMO-2013/10/E/ST3/00215 and no. UMO-2015/16/T/ST3/00506 and of Polish Ministry of Science and Higher Education under the project no. IP2014040473 are acknowledged.

\begin{suppinfo}
\beginsupplement

\begin{figure}
  \includegraphics[width=1\linewidth]{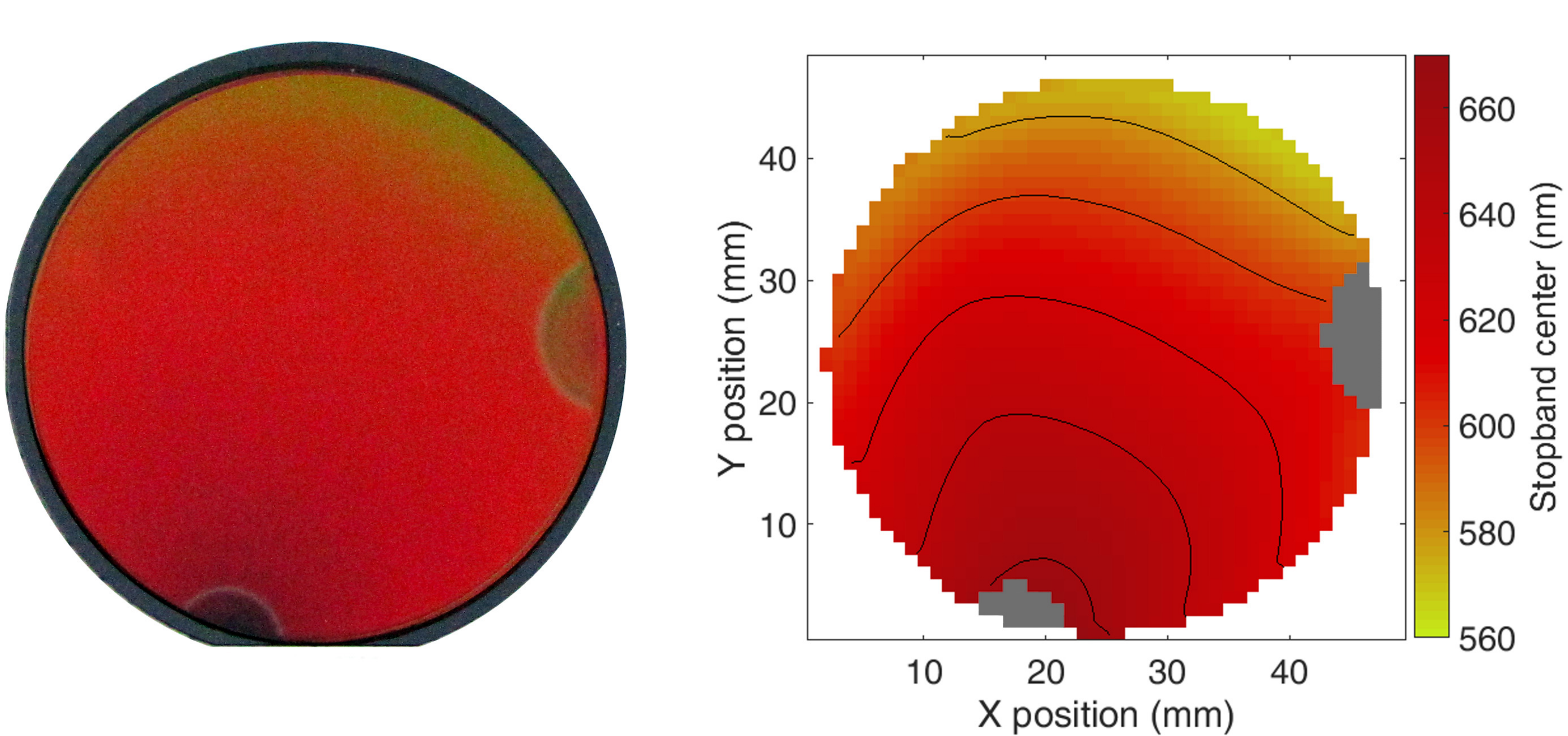}
\caption{a) A real color photograph, b) a color map showing a central wavelength of the stopband determined from the reflectivity mapping performed on the 12 DBR sample.
}
\label{fig:SIstopband_center}
\end{figure}

The Suppl. Fig.~\ref{fig:SIstopband_center}a) shows the real photograph of the 12 DBR sample as well as a color map representing the spectral position of the stopband centre determined from the reflectivity mapping. It is seen that the both images are mutually consistent.

The difference of the eigenmodes energy as a function of the position on the sample, thus of the detuning, for 12 DBR and 6 DBR sample is shown in Fig.~\ref{fig:17vs45}a) and Fig.~\ref{fig:17vs45}b), respectively. In agreement with the expectation and the results of our simulations, the eigenmodes splitting at resonance is larger when the cavities separation is smaller.
\begin{figure}
  \includegraphics[width=0.65\linewidth]{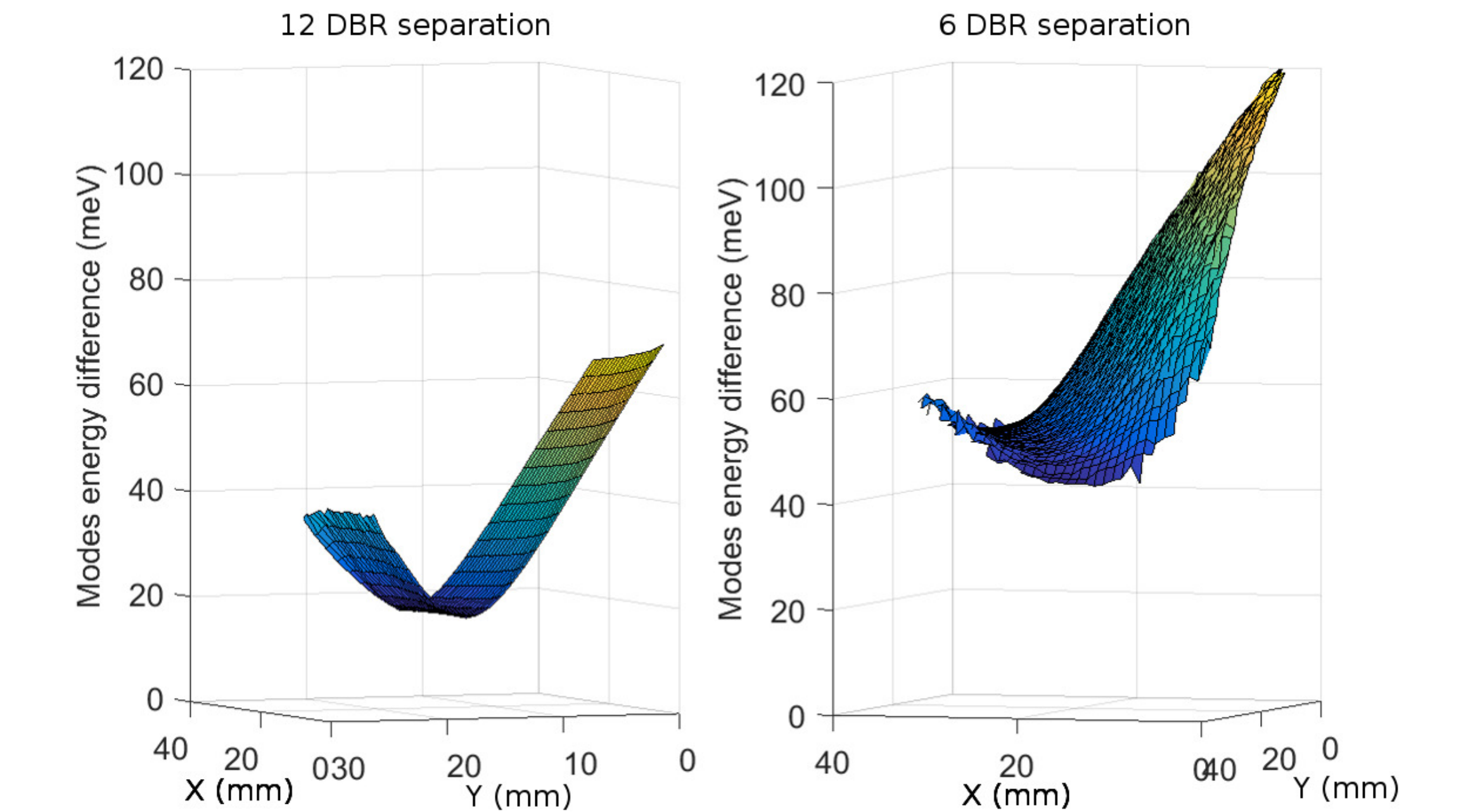}
\caption{Difference of the eigenmodes energy as a function of the position on the sample for coupled ZnTe planar microcavities separated by a) 12 and b) 6 DBR pairs.}
  \label{fig:17vs45}
\end{figure}

\end{suppinfo}

\bibliography{MS_coupled_cavities_bibliography}

\begin{mcitethebibliography}{42}
\bibitem[com()]{comment_kappa}
The values of the $\kappa$ obtained from the calculation are negative. We
  prefer, however, to plot the absolute value of the $\kappa$ in Figure 1,
  since such presentation emphasizes better that the cavities interaction
  strength (represented by the $\kappa$) decreases with the cavities
  separation.\relax
\mciteBstWouldAddEndPunctfalse
\mciteSetBstMidEndSepPunct{\mcitedefaultmidpunct}
{}{\mcitedefaultseppunct}\relax
\EndOfBibitem
\end{mcitethebibliography}

\end{document}